\documentstyle[preprint,pre,aps]{revtex}

\begin{document}

\title{\bf Magnetophoresis of Tagged Polymers}
\author{G. T. Barkema}
\address{Institute for Advanced Study, Olden Lane, Princeton NJ 08540, USA}

\author{G. M. Sch\"utz}
\address{Theoretical Physics, University of Oxford,\\
1 Keble Road, Oxford, OX1 3NP, United Kingdom.}

\maketitle

\begin{abstract}
We present quantitative results for the drift velocity of a polymer in a gel
if a force (e.g. through an electric or magnetic field) acts on a tag,
attached to one of its ends.  This is done by introducing a modification
of the Rubinstein-Duke model for electrophoresis of DNA. We analyze this
modified model with exact and Monte Carlo calculations. Tagged magnetophoresis
does not show band collapse, a phenomenon that limits the applicability of
traditional electrophoresis to short polymers.
\end{abstract}


If electrophoresis is applied for the separation of large polymers, for
instance fragments of DNA, the velocity becomes independent of the length.
This phenomenon is called ``band collapse'' and is widely known amongst
experimentalists.  A consequence is that long polymers cannot be separated
efficiently as they will all travel with approximately the same velocity.
Surprisingly, these experimental results can be predicted from a quite
simple one-dimensional lattice gas model, proposed by Rubinstein
\cite{rubinstein} as a model for the dynamics of entangled polymers
(the {\it repton} model), and adapted by Duke as a model for gel
electrophoresis of DNA \cite{duke}. It has been studied both analytically
\cite{rubinstein,widom,kvanl}, and by Monte Carlo simulations
\cite{duke,widom,barkema}. The same lattice gas model, but seen in a
different context, is of interest in other areas: it describes coarsening
\cite{rama}, diffusion of hard-core particles \cite{Ligget} with open
boundary conditions, and interface growth \cite{KS}.

In this work we focus on its application to the motion of long polymers in a 
gel. We study a modification of the traditional electrophoresis:
instead of pulling uniformly charged polymers with an electric field,
we propose to pull on one of the ends of the polymer,
which can be realized by attaching a magnetic tag to the polymer and
applying a magnetic field. We will show that
according to the repton model, in this case the drift velocity stays
length-dependent, allowing for the efficient separation of long polymers.
In previous work on the application of tags in electrophoresis
\cite{ulanovsky}, a force acts on the polymer and not on the tag, whereas
we propose to pull on the tag and not on the polymer.

First we will give a short description of the repton model, list
some of its properties, and compare the properties of the repton model
with experimental results on DNA electrophoresis in agarose gels.
Next, we will use the repton model to study what happens if 
a force pulls only on one end of the polymer rather than uniformly along the
whole molecule. Finally, we will outline some possible experimental
realizations and summarize our results. 

The repton model is illustrated in Fig. 1. In this model,
the gel is represented as a collection of pores that are arranged in a
square lattice. A polymer, represented as a chain of $N$ polymer segments
or {\it reptons} (the black dots in Fig. 1) moves through
this gel.

These reptons move from pore to pore diagonally according
to the following rules:

\noindent
(a) reptons in the interior of the chain move only along the sequence
of pores occupied by the chain. This restriction ensures that the only
possible mechanism of movement is {\it reptation}.

\noindent
(b) at least one repton must remain in each pore along the chain but
otherwise the number of reptons in a pore is unrestricted. By imposing
this rule we achieve that the polymer has some elasticity but does not
stretch to infinite length.

\noindent
(c) the two chain end segments may move to adjacent pores provided that
the rule (b) is not violated.

The dynamics of the model are defined as follows: Each possible move
(each repton moving in any direction) is tried with a constant rate,
setting the unit of time. This represents the diffusive motion of the
polymer. To make it also a model for electrophoresis, Duke has
introduced a small modification to this original repton model. To each
repton a negative electric charge is assigned on which the electric field
$E$ acts, breaking the symmetry.  Instead of unit rates for the allowed
moves in the positive and negative $x$-direction, the rate for allowed
moves in the positive $x$-direction (direction opposite to the electric
field) equals $\exp(+E/2)$, while the rate for allowed moves in the negative
$x$-direction equals $\exp(-E/2)$.

For the application to electrophoresis the quantity of interest is the
velocity of the polymer as a function of the field strength and the length
of the polymer.
In extensive Monte Carlo simulations \cite{barkema} it was found that a
data collapse could be obtained if the rescaled velocity $vN^2$ is plotted
as a function of the rescaled electric field $NE$. Plotted in such a
way, all data points were found to be located on one single curve, given by
\begin{equation}
\label{vcollapse}
\frac{vN^2}{\beta} =
\left[
\left( \frac{NE}{\alpha} \right)^2+
\left( \frac{NE}{\alpha} \right)^4
\right]^{1/2},
\end{equation}
where $\beta=5/6$ and, as expected from the Nernst-Einstein relation,
$\alpha/\beta=DN^2=1/3$ where $D$ is the diffusion coefficient of the
polymers.
For small $NE$, the rescaled velocity $vN^2$ increases linearly with $NE$.
corresponding to $v=E/3N$. For large $NE$, the rescaled velocity shows
a quadratic behavior. This corresponds to $v \sim E^2$.

It turns out that a collapse of experimental data for DNA electrophoresis
in agarose can be achieved with Eq. \ref{vcollapse}, as illustrated in
Fig. 2 for data points published by Heller et al. \cite{heller}
and rearranged as in \cite{barkema2}.
Taking the velocity $v$ in cm/h, the electric field $E$ in V/cm, and
DNA lengths $N$ in kilobase-pair (kb), one gets
$\alpha/\beta=(0.5 \pm 0.1) p^{3/2}$ and $\beta=(17 \pm5) p^{-5/2}$,
where $p$ is the gel concentration in percent agarose by weight.

This surprisingly good agreement of experimental data with the theoretical
prediction (\ref{vcollapse}) indicates that the repton model contains all
important ingredients for DNA electrophoresis: entanglement of the polymer
in the gel, entropic elasticity of the polymer, and reptation as the dominant
mechanism of movement.

In this work we use the repton model to study the behavior of polymers
in a gel if a force is applied only to one of the
ends of the polymers. To do this within the framework of the
repton model, we preserve the rules that specify which moves are allowed
and which ones are not. We assume that the force acts only on repton 1.
The attempt frequencies are now not uniform along the chain:
as in the repton model for DNA electrophoresis as adapted by Duke, repton 1
moves in the positive and negative $x$ direction with rates $q=\exp(E/2)$ and
$q^{-1}=\exp(-E/2)$, respectively;
as in the original repton model of Rubinstein, all allowed moves of all
other reptons occur with unit rate.

These dynamical rules may be conveniently rephrased in lattice gas language
by writing a master equation for a one-dimensional exclusion process,
where the links of the repton chain are mapped to the sites of a
one-dimensional lattice with $L=N-1$ sites.
For each link $i=1 \dots L$ (connection between reptons $i$ and $i+1$) we
distinguish three cases: the link has a negative, a zero, or a
positive component in the $x$-direction. We identify these three cases with
the presence of an $A$-particle, a vacancy, and a $B$-particle, respectively,
on site $i$ of the lattice.
By this mapping the information on the motion of the polymer perpendicular
to the field gets lost, but this motion is purely diffusive and of no interest
for the study of the drift velocity of the polymer in direction of the field.
The drift velocity is in particle language given by the difference between
the current of $A$-particles and the current of $B$-particles along the chain.

In terms of particles, the dynamics may be described as follows:
If one site of a pair of nearest-neighbor sites is occupied by a particle
while the other site is empty, this particle hops to the empty site with
rate 1. In addition to these hopping events, particles can be
created and annihilated at the ends. At site $L$, $A$- and $B$-particles are
created and annihilated with rate 1, while at site 1 $A$-particles are
created and annihilated with rates $q$ and $q^{-1}$ and $B$-particles
with rates $q^{-1}$ and $q$, respectively.

By writing this process in a quantum Hamilton formalism,
the stochastic time evolution of the system is given by the Hamiltonian
$H = b_1(q) + b_L(1) + \sum_{i=1}^{L-1} u_i(1)$ of a spin-1 quantum chain
where $b_i(q) = q(1-n_i^A-a_i^{+}-b_i) + q^{-1} (1-n_i^B-a_i-b_i^{+})$, and
$u_i(q)=q(n_i^A n_{i+1}^0 + n_i^0 n_{i+1}^B-a_i a_{i+1}^{+}-b_i^{+} b_{i+1} )
+ q^{-1} (n_i^0 n_{i+1}^A + n_i^B n_{i+1}^0-a_i^{+} a_{i+1}-b_i b_{i+1}^{+})$.
Here $n_i^A \equiv E_i^{11}$, $n_i^B \equiv E_i^{33}$ and
$n_i^0 = 1-n_i^A-n_i^B \equiv E_i^{22}$ are projection operators on states
with spin $S^z = 1,0,-1$ resp. on site $i$, identified with an $A$-particle,
vacancy and $B$-particle resp. The operators
$a_i \equiv E_i^{21}$, $a_i^{+} \equiv E_i^{12}$, $b_i \equiv E_i^{23}$,
and $b_i^{+} \equiv E_i^{32}$ are annihilation and 
creation operators for $A$- and $B$ particles. $E_i^{jk}$ is the $3\times 3$
matrix with matrix elements $(E_i^{jk})_{\alpha,\beta}=\delta_{j,\alpha}
\delta_{k,\beta}$ acting on site $i$.

The average densities $\langle n_i^{A,B} \rangle$ at time $t$ satisfy
the continuity equations $\frac{d}{dt} \langle n_i^{A,B} \rangle = 
- \langle n_i^{A,B} H \rangle = \langle j_{i-1}^{A,B} \rangle -
\langle j_i^{A,B} \rangle $
with the particle currents 
\begin{eqnarray}
\label{2}
\langle j_i^{A,B} \rangle & = & \langle n_i^{A,B} n_{i+1}^0 \rangle
 - \langle n_i^{0} n_{i+1}^{A,B} \rangle \;\; 1 \leq i \leq L-1 \nonumber\\
\label{3}
\langle j_L^{A,B} \rangle & = & 
\langle n_L^{A,B} \rangle - \langle  n_{L}^0 \rangle \nonumber\\
\label{4}
\langle j_0^A \rangle & = & q \langle n_1^0 \rangle - 
q^{-1} \langle  n_1^A \rangle \nonumber\\
\label{5}
\langle j_0^B \rangle & = & q^{-1} \langle n_1^0 \rangle - 
q \langle  n_1^B \rangle \nonumber 
\end{eqnarray}

Among the quantities of interest are particle densities and correlations
in the steady state of the system, in particular the (space-independent)
stationary particle currents
$\langle j_i^{A,B} \rangle \equiv j^{A,B}$ which give information on the
internal structure
of the polymer and which give its average drift velocity $v=j^A - j^B$. 
Defining ${\cal J} = j^A + j^B$ and $m_i= n_i^A - n_i^B$,
stationarity of the probability distribution leads to the following relations: 
\begin{eqnarray}
\label{6}
{\cal J} & = & \langle n_{i+1}^0 \rangle - \langle n_i^0 \rangle \;\;
\hskip 0.5in (1 \leq i \leq L-1) \\
\label{7}
 & = & \frac{q+q^{-1}}{2}\left( 3 \langle n_1^0 \rangle -1 \right)
+ \frac{q-q^{-1}}{2} \langle m_1 \rangle \\
\label{8}
 & = & 1 - 3 \langle n_L^0 \rangle \\
\label{9}
v & = & \langle m_i n_{i+1}^0 \rangle - \langle n_i^0 m_{i+1} \rangle
 \;\; \hskip 0.2in (1 \leq i \leq L-1) \\
\label{10}
 & = & \frac{q-q^{-1}}{2}\left( 1 + \langle n_1^0 \rangle \right)
- \frac{q+q^{-1}}{2} \langle m_1 \rangle \\
\label{11}
 & = & \langle m_L \rangle
\end{eqnarray}
(\ref{6}) and (\ref{8}) give the exact average density of vacant sites
at site $i$ in terms of ${\cal J}$
\begin{equation}
\label{12}
\langle n_i^0 \rangle = \frac{1}{3}\left[1-(3L+1){\cal J}\right] + {\cal J} i
\end{equation}
and therefore the probability $1-\langle n_i^0 \rangle$ of finding only one
repton in the pore occupied by the $i^{th}$ repton. ${\cal L}=L-\sum_{i=1}^{L}
\langle n_i^0 \rangle = 2L/3 + {\cal J} L(3L-1)/6$ gives the average length of 
the polymer in terms of ${\cal J}$.
Eqs. (\ref{7}), (\ref{10}) and (\ref{12}) (for $i=1$) give 3 linear 
equations for the four quantities $v, {\cal J},\langle m_1 \rangle,
\langle n_1^0 \rangle$ relating the currents to the average densities at 
site 1. This may be used to derive an exact expression for $v$ in terms of
any of the three other expectation values. 

Now we study $v$ as a function of $q$ and $N=L+1$.  
The diffusion coefficient $D$ of the repton chains has not changed from
the original model, and is to lowest order in $1/N$ given by $D=1/(3N^2)$
\cite{praehofer}. The force on the chain is now independent of its length
and proportional to $E$. Combining this with the Nernst-Einstein relation
$v \approx F \cdot D$ yields $v = E/3N^2$ for small forces that do not disturb
the equilibrium configuration of the chain.

For intermediate fields and short chains, we can obtain the drift
velocity numerically exactly by computing the eigenvector with eigenvalue
0 of the quantum Hamiltonian $H$. Results for chains with up to 7 reptons
($L=6$) are presented in Fig. 3.
For intermediate fields and longer chains, we study the model by
Monte Carlo simulations. We performed these
simulations by means of a fast ``multi-spin'' algorithm as described in
an earlier paper~\cite{barkema}. Results are included in Fig. 3.

In the limit of an infinitely strong force, we can solve the model exactly
for any $L$. For an infinite creation rate of $A$-particles and
correspondingly infinite annihilation rate of $B$ particles on site 1
one has $\langle n_1^A \rangle =1$ and $\langle n_1^B \rangle =
\langle n_1^0 \rangle =0$. In this limit the stochastic process becomes
equivalent to a process on a chain on $L'=L-1$ sites where $A$ particles 
are created (annihilated) with rate 1 (0) on site 2 and where no $B$-particles
are created or annihilated. Going through the same calculations
as above gives expressions similar to (\ref{6}-\ref{11}), but with
(\ref{7}) and (\ref{10}) replaced by ${\cal J}=v=\langle n_2^0 \rangle$
and the range of definition in (\ref{6}) and (\ref{9}) replaced by
$2 \leq i \leq L-1$. Together with (\ref{12}) (for $i=2$) this shows that the
velocity approaches a constant that
depends on the length of the chain: $v = 1/(3L-2)=1/(3N-5)$. 
This asymptotic regime starts around $E=3$ as can be seen from the exact
and Monte Carlo results shown in Fig. 3.

Thus, according to this
modified repton model, there is no band collapse if we pull on one end
of the chain, even in the limit of an infinitely strong field.
In this limit, the polymer is not fully stretched but has an average
length ${\cal L}=L (5+(3L-2)^{-1})/6 \approx 5L/6$.

Experimentally, a realization of pulling only at one end of each
polymer, is to attach to the end a magnetic bead.
Magnetic beads are produced commercially; for a discussion of their
properties and traditional application in biotechnology we refer to
\cite{haukanes}.
These commercially available magnetic beads are rather large for use in
a normal agarose gel, and have the risk to get stuck, but a dense
polymer solution may be a suitable replacement for the agarose gel. 
It is interesting to point out that according to our model, variations
in the force on the end of the polymer, (e.g. due to differences in
the magnetic beads) do not
lead to a significant band widening if the force is sufficiently large,
as in that case the drift velocity is independent of the force (see figure 3).

If such ``magnetophoresis'' turns out to be possible, this experimental
approach might also avoid band collapse in RNA electrophoresis and make
sequencing of longer fragments possible.
Another interesting topic is the behavior of a polymer that experiences a
magnetic force on one end, in addition to a uniform electric force on the
whole chain.

Considerable effort has been invested into solving the Rubinstein-Duke
model for electrophoresis, but to the best of our knowledge no
rigorous results have been obtained, except for very short polymers 
\cite{widom} and for a related model with periodic boundary conditions
\cite{kvanl}. In zero field the model is equal to the symmetric exclusion
process with tagged particles, which is (at least) partially integrable
by the matrix Bethe ansatz discussed in \cite{stsc}.
An open question is whether open boundary conditions as proposed here
leave the system completely integrable; this would allow for a full exact
solution of both the model proposed here and the original Rubinstein model.

To conclude: previous work has shown that the Rubinstein-Duke model for
DNA electrophoresis provides a good description of experimental results of 
electrophoresis of DNA fragments in agarose gels~\cite{barkema2}.
We propose a modification in the model to describe pulling only on the end of
the polymer instead of uniformly along the chain. With Monte Carlo simulations
and exact calculations we predict that this provides a way
of avoiding band collapse, a phenomenon that severely limits the separation 
of long polymers by traditional electrophoresis.
Experimental realizations of this modified model and various
applications are discussed.

We thank J.-L. Viovy, C. Heller and S. Margel for interesting comments on
experimental realizations.
G.M.S. would like to thank C. Godreche for stimulating discussions.
This work was supported by the EPSRC under Grant No. GR/J78044 (G.T.B.),
and under the Human Capital and Mobility program of the European Community
(G.M.S.).

\bibliographystyle{prsty}

\begin{figure}
\label{fig1}
\caption{Left: Two-dimensional representation of the repton model of a
long polymer. The reptons (black circles) are located in the pores of
the gel (squares of the lattice).
Right: projected repton model where the reptons are numbered along the
chain, and the $x$-coordinate of repton $i$ is plotted as a function of $i$.
The arrows indicate the allowed moves.}
\end{figure}

\begin{figure}
\label{fig2}
\caption{Experimental data on the rescaled drift velocity $vN^2/\beta$
as a function of the rescaled electric field $NE/\alpha$. The solid
curve is the theoretical prediction by the repton model, see eq.
\protect\ref{vcollapse}. The data points are obtained from
\protect\cite{heller} and rearranged as in \protect\cite{barkema2}. }
\end{figure}

\begin{figure}
\label{fig3}
\caption{Drift velocity $v$ as a function of the force $E$. The solid
lines are exact results for $L=2 \dots 6$, the dashed lines are obtained
from Monte Carlo simulations for $L$=7,10,15,20,30 and 50. Statistical
errors smaller than 2\% are omitted.}
\end{figure}

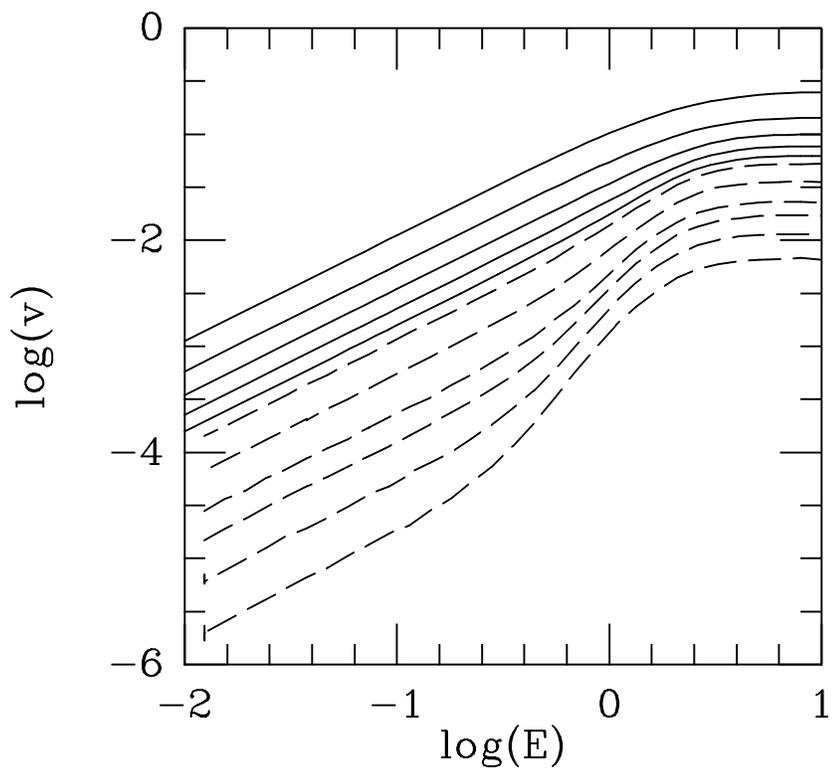
\end{document}